# Stable beam operation of approximately 1 mA beam under highly efficient energy recovery conditions at compact energy-recovery linac

Hiroshi Sakai[#], Dai Arakawa, Takaaki Furuya, Kaiichi Haga, Masayuki Hagiwara[*], Kentaro Harada, Yosuke Honda, Teruya Honma, Eiji Kako, Ryukou Kato, Yuuji Kojima, Taro Konomi[**], Hiroshi Matsumura, Taichi Miura, Takako Miura, Shinya Nagahashi, Hirotaka Nakai, Norio Nakamura, Kota Nakanishi, Kazuyuki Nigorikawa, Takashi Nogami, Takashi Obina, Feng Qiu[***], Hidenori Sagehashi, Shogo Sakanaka, Miho Shimada, Mikito Tadano, Takeshi Takahashi, Ryota Takai, Olga Tanaka, Yasunori Tanimoto, Akihiro Toyoda, Takashi Uchiyama, Kensei Umemori, Masahiro Yamamoto, Go Yoshida
High Energy Accelerator Research Organization (KEK), Tsukuba, Ibaraki, Japan

Nobuyuki Nishimori[*]
National Institutes for Quantum and Radiological Science and Technology (QST), Sayo-cho, Hyogo, Japan

Ryoichi Hajima[****], Ryoji Nagai[*****], Masaru Sawamura[****]
National Institutes for Quantum and Radiological Science and Technology (QST), Tokai, Ibaraki, Japan

*Abstract*

A compact energy-recovery linac (cERL) has been under construction at KEK since 2009 to develop key technologies for the energy-recovery linac. The cERL began operating in 2013 to create a high-current beam with a low-emittance beam with stable continuous wave (CW) superconducting cavities. Owing to the development of critical components, such as the DC gun, superconducting cavities, and the design of ideal beam transport optics, we have successfully established approximately 1 mA stable CW operation with a small beam emittance and extremely small beam loss. This study presents the details of our key technologies and experimental results for achieving 100% energy recovery operation with extremely small beam loss during a stable, approximately 1 mA CW beam operation.

## I INTRODUCTION

The energy-recovery linac (ERL) can accelerate high-current beams with low emittances and short bunches. The ERL concept was proposed by M. Tigner in 1965 [1], based on the following principles: The high-current beam generated by the high-brightness electron gun is first accelerated by the accelerating cavities and recirculated. The beam returns to the same cavities and is decelerated by the same accelerating cavities. This beam energy, by decelerating at the same cavities, is stored in the same accelerating cavities and "reused" for the acceleration of the beam from injector. This energy-recycling scheme is a key feature of the ERL. In particular, using a superconducting cavity with no cavity-wall losses for accelerating cavities enables the approximately 100% energy recovery operation of high-current CW beams. Using this ERL scheme, low-emittance, high-current, and short-bunch beams can be produced; however, this cannot be achieved using a conventional storage ring. Furthermore, the technology is sustainable and can significantly reduce the operating power by reusing the beam power in superconducting cavities.

ERL applications vary widely. For example, using ERL for collider accelerators in high-energy experiments [2] and nuclear particle experiments [3, 4] has been proposed. Furthermore, extremely high-intensity FEL light sources, which are not possible with existing light sources, have been proposed [5, 6]. In addition, ERL-based compact X-ray sources using laser Compton scattering [7] and THz light source applications have been proposed [8]. The key technologies for these applications mentioned above include a high-brightness electron gun with low emittance and high current, a superconducting accelerating cavity for energy recovery, and a beam-handling technology for energy recovery with extremely small beam loss.

Over the years, various test facilities have been constructed worldwide to improve the energy recovery technologies for high-brightness, high-current beams. In the 2000s, the Japan Atomic Energy Agency [9] in Japan and the Jefferson Laboratory (Jlab) [10] in the U.S. conducted energy recovery using superconducting cavities in infrared free-electron laser (FEL) experiments around when the superconducting technology matured. For example, Jlab achieved a maximum energy recovery of 9 mA at 150 MeV. During its operation, the beam instability of the HOM-beam-break-up (BBU) appeared from a few mA. The effect of the beam emittance on the HOM-BBU instability was observed. Although stable beam operation has been a challenge for higher-current CW beam operations with superconducting cavities, ERL operation has been performed using energy recovery with normal-conducting cavities in a recuperator [11], where current of 30 mA was recovered under energy recovery conditions. However, the wall loss of the cavity owing to normal conduction was very large, and the acceleration gradient was very low. Notably, a multiturn ERL is a more attractive operation style for saving operation and construction costs owing to the reduced superconducting cavity section. Some facilities successfully operate with a low-current beam for multi-turn ERL operations [4,12]. High-brightness electron guns have been developed at CBETA, the ERL test machine at Cornell University, and have achieved an emittance of less than 1 mm mrad with a 390 kV DC electron gun at a high beam current of 65 mA [13]. Consequently, the energy recovery of the high-current beam is the next issue that needs to be addressed. In particular, the stable energy recovery operation of a high current without beam loss with high gradients in a superconducting cavity is of utmost importance for developing ERLs worldwide. For example, radiation due to beam loss during the energy recovery operation causes damage and/or heat load to the components of the beam line. Furthermore, uncontrolled beam loss is a severe problem in the design of radiation shields, which are used for protection during beam operations. Under these conditions, a compact ERL (cERL) was constructed to ensure stable high-current ERL beam operation.

___________________________________
[#] Corresponding author: hiroshi.sakai.phys@kek.jp
[*] Present address: National Institutes for Quantum Science and Technology (QST), Sendai, Miyagi, Japan
[**] Present address: Michigan State University, East Lansing, MI, USA
[***] Present address: Institute of Modern Physics (IMP), China
[****] Present address: National Institutes for Quantum Science and Technology (QST), Kizugawa, Kyoto, Japan
[*****] Present address: National Institutes for Quantum Science and Technology (QST), Chiba, Chiba, Japan.

The remainder of this paper is as follows: Section II briefly explains the cERL. The beam-tuning methods for reducing the beam loss of a CW high-current beam under energy recovery are also introduced, explaining the important components of the cERL in Section II. Section III presents the measurement results of the beam parameters and energy recovery efficiency under approximately 1 mA stable beam operation. Section IV discusses the correlation between the radiation dose outside the radiation shield during energy recovery and the energy recovery efficiency and the beam parameters under energy recovery conditions. Section V concludes the study.

## II MATERIALS AND METHODS

This section presents a brief overview of the cERL, reviewing the critical components for high-current beam operation under energy recovery conditions. It also presents a method for high-current beam tuning of approximately 1 mA under energy-recovery conditions.

### II-1 Experimental setup of cERL

#### II-1-1 Overview of compact ERL

At the High Energy Accelerator Research Organization (KEK), the cERL was constructed in 2009 to study the feasibility of a future 3 GeV ERL light source [14]. Fig. 1 shows a schematic of the cERL. The cERL comprises a high-brightness photocathode DC gun [15] capable of producing ultra-low emittance beams at a high average current for long periods, a green 1 W power laser for irradiating the photocathode, an injector cryomodule with three two-cell 1.3 GHz superconducting cavities [16], a main-linac superconducting cavity with two nine-cell 1.3 GHz superconducting cavities, where energy recovery is performed [17], a recirculation loop for energy recovery to maintain high-beam quality, and a beam dump for the decelerated beam. A main-linac superconducting cavity was designed to suppress the HOM-BBU, which made the beam unstable in the Jlab-FEL for high-current beam operation [10], for a beam energy recovery condition of more than 600 mA [18]. The details of the cERL have already been described in Ref. [14], and energy recovery has been achieved with a 10-µA CW beam since commissioning only began in 2013.

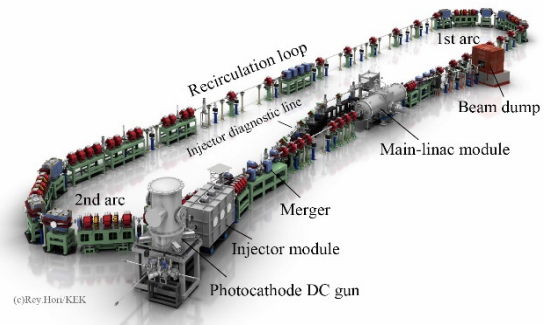

Figure 1: Schematic view of the cERL [14].

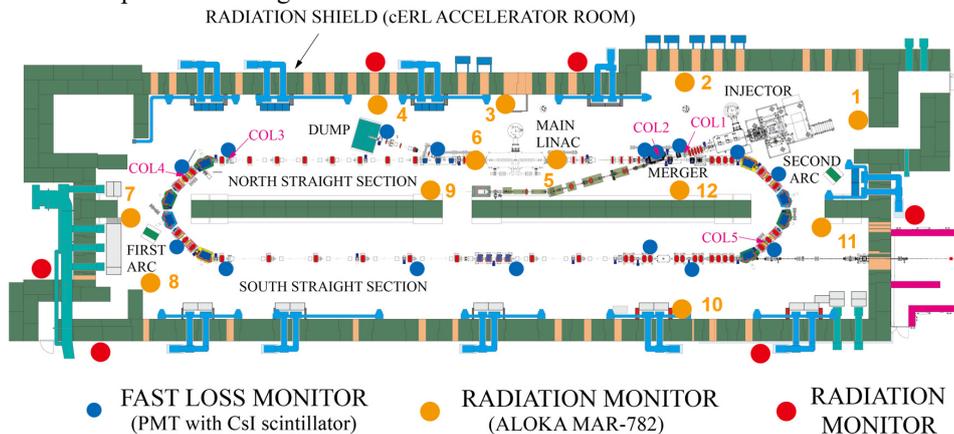

Figure 2: Setting of local loss monitor and radiation monitor (blue circles). The locations of the local collimators are expressed as COL1,2,3,4 and COL5. ALOKA radiation monitors (yellow circles) are also shown with numbering. [14]

**II-1-2 Critical components for CW beam operation**

In this section, we describe the critical components for CW beam operation at 1 mA. Fig. 2 shows the settings of the radiation monitor and collimator system for the high-power CW beam operation. For high-current CW beam operation under highly efficient energy recovery conditions, the small beam loss along all beamlines must be reduced. We installed five beam collimators to achieve such low beam losses, as shown in Fig. 2. Each collimator had four water-cooled copper rods, which were inserted in both horizontal and vertical directions, as shown in Fig. 3. The beam halo, defined as a collection of low-density particles around the core of the transverse beam distribution, and the beam tail, defined as a collection of low-density particles around the core of the longitudinal beam distribution, are primarily eliminated by two collimators, COL1 and COL2, located in the low-energy (maximum of 5 MeV) section. The remaining beam halos were eliminated by the three other collimators: COL3–COL5. Each collimator was locally shielded with a 20-mm-thick lead jacket and lead blocks. The tolerable amounts of beam losses were 1 μA and 10 μA (at 6 MeV) for the local shields at the COL1 and COL2 collimators, respectively, and 50–100 nA (at 26 MeV) at the COL3–COL5 collimators. We prepared a local fast beam loss monitor to detect the beam loss and cut the drive laser of the electron gun for safety. This monitor comprises a photomultiplier tube (PMT Hamamatsu, R11558) and a CsI (pure) scintillator with a size of 25 mm × 10 mm × 10 mm to observe the local beam loss and interlock with the beam within 10 μs. We prepared 16 local fast beam loss monitors in the cERL, denoted by the solid blue circles in Fig. 2. These loss monitors were placed approximately 10 cm far from the cERL beamline.

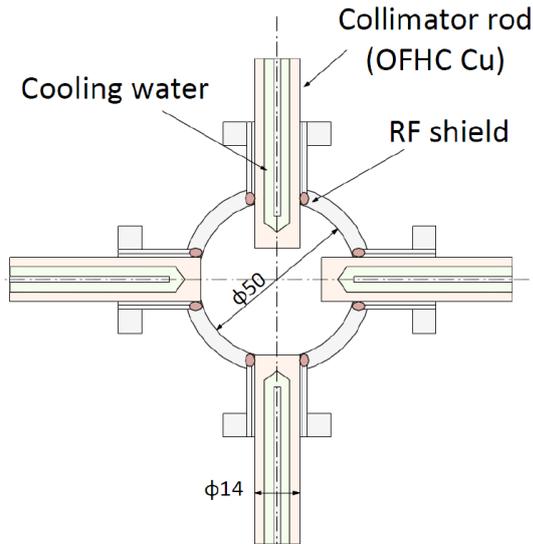

Figure 3: Cross-section of collimators (COL1 and COL2). The upper and lower collimators and the left and right collimators are set so that they do not collide with each other.

Our cERL was shielded with a 1-m concrete block for the roof and a 1.5-m concrete block for the side wall to reduce the radiation outside the concrete shield. The 12 radiation monitors (Hitachi Aloka Medical, MAR-782) (referred to as the "Aloka radiation monitor" hereafter) were located in a cERL accelerator denoted as the solid yellow circles, as shown in Fig. 2. Furthermore, we directly measured the radiation on the cERL roof while walking on the roof.

The beam was rastered at high speed by a fast-steering magnet just before the beam dump to avoid local heating due to a high-current beam on the beam dump.

*II-2 Method of beam tuning toward CW 1 mA beam under energy recovery conditions*

**II-2-1 Optics and beam-tuning method**

For beam tuning, we first used the burst mode before the CW beam operation. In this burst mode, burst lengths of 0.1–1.2 μs, and a burst repetition rate of 5 Hz were used. This short burst length enabled beam profile monitoring. The bunch charge was typically 0.7 pC at 1.3 GHz beam repetition. Beam-optics matching was performed upstream from the electron gun [19]. The energy of the injector part was adjusted by fine-tuning the amplitude and phase of the injector cavity; it was determined by the bending magnet of the merger at the injection section to the circumference section, as shown in Fig. 4. The energy of recirculation is as follows: First, the beam energy was optimized to set the design voltage of ML1, as shown in Fig. 4, by scanning the phase to the on-crest condition. Second, we set the design voltage of ML2, as shown in Fig. 4, using the same method as that for ML1. For adjustments of both ML1 and ML2, the beam energy was measured using the position of the screen monitor immediately after the bending magnet at the entrance of the first arc section to set the on-crest condition. After tuning the beam optics in the recirculation loop without bunch compression in the arc section, the beam was decelerated using ML1 and ML2 to achieve energy recovery. After passing through ML1 and ML2, the beam energy was minimized by changing the path length of the circumference of the cERL. The path length was changed mainly by changing the beam optics at the second arc section and the chicane magnets, as shown in Fig. 4. The beam energy after deceleration was measured using a screen monitor immediately after bending the merger section at the dump line. The bending angle of the bending magnet at the merger at the dump line was set such that the beam energy at the dump line was the same as that of the injector section when the main linac cavities of ML1 and ML2 were detuned and the beam directly passed through the dump line without the recirculation loop.

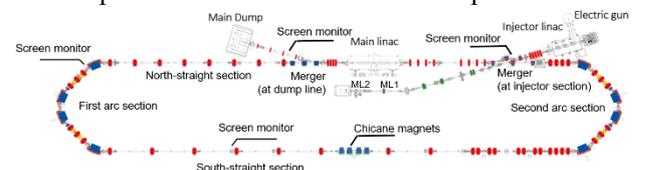

Figure 4: The beamline configuration on the cERL.

## II-2-2 Reduction of unexpected beam halo by using collimators for CW beam operation of 1 mA

The beam was produced by directly using a DC gun and circulated once along the recirculation loop. When the produced beam has an unwanted halo or tail, it is also transported, resulting in beam loss in this recirculation loop. In particular, owing to the time response of the cathode material, the beam tail is produced after a short laser pulse and forms a beam halo in the transverse direction because the tail experiences different accelerating voltages compared to the core of the beam [20].

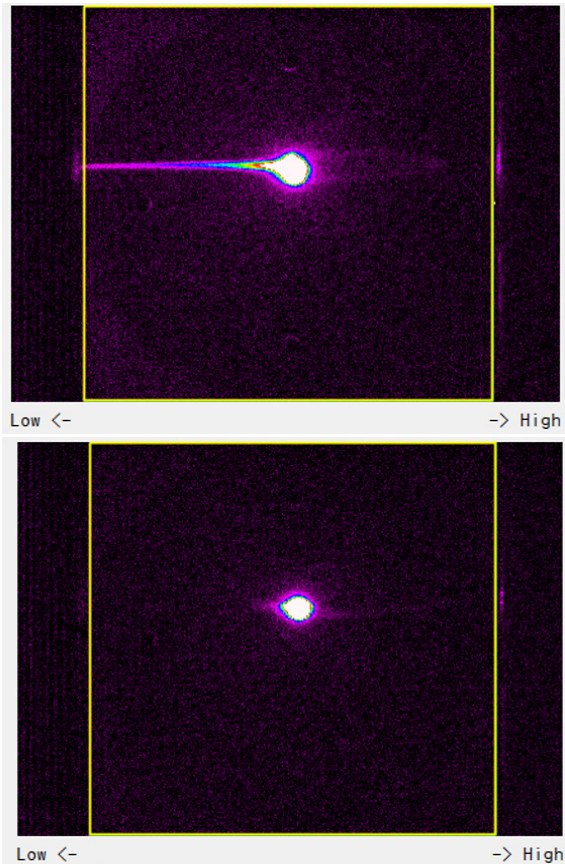

Figure 5: (Top) Beam profile by screen monitor, where the dispersion is approximately 0.23 m after the COL1 position just before entering the merger section and before inserting the collimator (COL1). (Bottom) Beam profile by the same screen monitor after inserting collimator (COL1). In both figures, the beam intensities in the core of the measured beam profile are saturated to emphasize the beam halo.

The beam halo and tail were cut by the collimator before the main linac acceleration to reduce radiation loss. Therefore, we used COL1 and COL2 to cut the beam halo and tail. The tail of COL2 can be cut off owing to the finite dispersion function at the COL2 position. However, to cut the tail more efficiently at the COL1 positions, which have no dispersion function, the orbit of the injector cavity was intentionally threaded at a finite angle so that the tail was kicked by the field of the injector cavity. Fig. 5 shows an example of beam collimation. The beam profiles were measured at the position where the dispersion was approximately 0.23 m after the COL1 position. The low-energy tail part of the beam was successfully reduced using the vertical collimator COL1, as shown in Fig. 5. Notably, we successfully cut the tail at the COL2 position. However, as shown in Fig. 5, the beam tail was kicked by the injector cavity field, and the longitudinal beam profile was projected onto the larger vertical transverse beam profile such that the vertical collimator in COL1 could cut the beam tail more effectively. We also conducted detailed, precise collimator tuning using a fast loss monitor while monitoring the beam current of the beam dump. Fig. 6 shows the typical tuning of the collimator at COL2. After setting a length of a few millimeters from the beam center, the measured signals of the fast-loss monitor reduced drastically, as shown in Fig. 6. We also monitored the beam current at the beam dump to avoid scraping the beam core during collimator tuning. COL3, COL4, and COL5 were also used to reduce beam loss, if necessary.

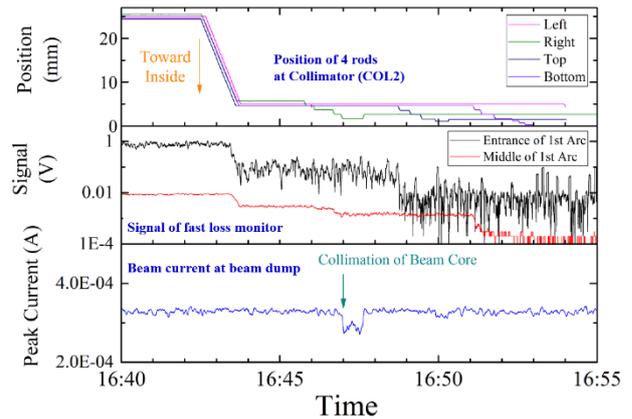

Figure 6: (Top) Position of four rods at a collimator (COL2). (Middle) Measured signal of a fast-loss monitor under collimator tuning. (Bottom) Beam current at beam dump under collimator tuning. The horizontal axis shows the elapsed time.

After sufficiently reducing the loss in burst mode, the operation was switched to CW mode. The laser intensity of the electron gun gradually increased, and the current increased while ensuring that the ALOKA radiation monitor did not exceed the threshold value. The BPM, vacuum, and power consumption of the main superconducting acceleration cavity during the operation were monitored. However, the beam current wobbled slightly. The operation was conducted at approximately 0.9 mA to prevent the current from exceeding 1 mA even for a moment. This beam was operated such that it hit the entire beam dump via the fast-steering magnet immediately before the beam dump.

.

# III EXPERIMENTAL RESULTS

## III-1 Parameters of energy recovery condition

Three main CW energy recovery runs of approximately 1 mA were performed under various conditions. Table 1 lists the parameters of the three CW beam energy recovery runs. Runs 1 and 3 are the energy recovery operation parameters when the total energy is 20 MeV and the injector energy is 2.9 MeV. In detail, Run 1 is operated at 1.3 GHz with 0.7 pC/bunch, while Run 3 is operated during an X-ray generation experiment in laser Compton scattering, with a repetition rate of 162.5 MHz and a high charge of 5.5 pC/bunch. Conversely, Run 2 shows the energy recovery operation parameters when the total energy is lowered to 17.5 MeV. For Run 2, the energy ratio in the merger section was 1:6, which is smaller than the energy ratio of 1:7 in Runs 1 and 3.

Table 1: Beam parameters for approximately CW 1 mA energy recovery operation

| Parameters | Run 1 | Run 2 | Run 3 |
|---|---|---|---|
| Beam repetition rate (MHz) | 1300 | 1300 | 162.5 |
| Bunch charge (pC) | 0.7 | 0.7 | 5.5 |
| Gun HV (kV) | 390 | 500 | 390 |
| Injector energy (MeV) | 2.9 | 3.0 | 2.9 |
| Energy of recirculation loop (MeV) | 20.0 | 17.6 | 20.0 |
| Momentum ratio between injector and recirculation loop | 1:7 | 1:6 | 1:7 |

## III-2 Measurement results of beam emittance and energy spread

Before the CW operation, we performed beam–optics matching, as described in Sec. II-2-1. The typical measurement results of the beam size after beam tuning and optics matching in Run 1 are shown in Fig. 7. The horizontal and vertical beam sizes measured at each screen monitor on the cERL beam line in Run 1 were roughly consistent with our setting optics by assuming the 0.3 mm mrad normalized emittances.

Table 2 lists the results of the normalized emittance measurements for Runs 1, 2, and 3. The normalized emittances were measured by scanning the quadrupole magnet in front of the monitor screen located at south-straight section, as shown in Fig. 4. These normalized emittances were measured after optical matching, before changing the CW energy recovery beam operation. For low-charge beam operation of 0.7 pC/bunch on Run 1 and 2, we achieved the lower emittance of 0.3–0.4 mm mrad. In the case of the high-charge operation of the 5.5 pC/bunch in Run 3, we also achieved a small beam emittance of 1–2 mm mrad in the cERL beam line.

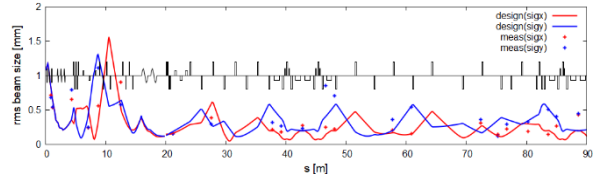

Figure 7: The beam optics of the cERL beam line from the electron gun to the 2nd arc section. The solid red (blue) line shows the horizontal (vertical) beam size calculated by our beam optics, under conditions of 0.77 pC/bunch and 0.3 mm mrad emittances. The dotted red (blue) points show the measured horizontal (vertical) beam size at each screen monitor on Run 1.

Table 2: Measured normalized emittances for three runs

| Measured emittance (mm mrad) | Run 1 | Run 2 | Run 3 |
|---|---|---|---|
| Horizontal at C | 0.259 ± 0.005 | 0.406 ± 0.008 | 2.11 ± 0.17 |
| Vertical at C | 0.267 ± 0.092 | 0.256 ± 0.014 | 1.09 ± 0.05 |

We also measured the energy spread using a screen monitor set at the entrance of the first arc section with a dispersion function of 0.49 m. The measured beam size on this screen monitor was 0.36 mm in r.m.s. From these measurements, we estimated an energy spread of less than 0.07% in r.m.s on all three runs.

## III-3 Measurement results of energy recovery efficiency

Energy recovery was conducted at the main linac to achieve the high-current beam operation of the recirculation loop, and no energy was recovered at the injector section. The most important performance of the energy-recovery linac is to recover energy without losing beam energy in the recirculating loop. Therefore, a high energy recovery efficiency of the main linac is important for obtaining high-power energy recovery linac without beam loss. The energy recovery efficiency of the main linac ($\varepsilon_{rf}$) is given by Eqs. (1) [21]:

$$\varepsilon_{rf} = (P_{rf,acc} - P_{rf,load}) / P_{rf,acc} \quad (1)$$

This is defined as $P_{rf,acc} = V_c \times I$. $V_c$ is the total accelerating voltage on the main linac, and I is the beam current through the main linac. $P_{rf,load}$ is the remaining power that is not recovered owing to beam loading. $P_{rf,load}$ is equal to $P_{rf,acc}$ when the energy is not recovered. Conversely, a 100% energy recovery efficiency of the main linac is obtained when beam loading in the ERL mode vanishes completely. This implies that the RF power to the main linac cavities should not change during energy recovery and should not depend on the beam current. If the energy recovery is not perfect, we can observe a difference in $P_{rf,load}$ with and without the beam during the energy recovery operation.

We can measure the change in $P_{rf,load}$ in each cavity with and without the beam during beam operation. $P_{rf,load}$ was expressed as the change in difference ($P_{in,cav}$-$P_{ref,cav}$), where $P_{in,cav}$ shows the input power to each main linac cavity and $P_{ref,cav}$ shows the reflected power from each main linac cavity. We measured the changes of ($P_{in,cav}$-$P_{ref,cav}$) of cavities of ML1 and ML2 when the beam current (I) increased approximately 1 mA relative to the beam current of 0 mA. Finally, $P_{rf,load}$ through the all main linac was estimated. The energy-recovery efficiency ($\varepsilon_{rf}$) in Eq. (1) is rewritten during the CW beam operation at the cERL as

$\varepsilon_{rf}$ (%) = (1 − ($P_{in}$-$P_{ref}$)/ $P_{rf,acc}$) x 100%  (2)

,where the beam power is expressed as $P_{rf,acc} = (V_{c1}+V_{c2})$ x I, with the ML1 cavity voltage ($V_{c1}$) and ML2 cavity ($V_{c2}$) and $P_{in}$ ($P_{ref}$) expressed as the sum of $P_{in,cav}$ ($P_{ref,cav}$) of ML1 and ML2 cavities, respectively. We could stably operate by monitoring the RF field using a pickup monitor with LLRF control [22].

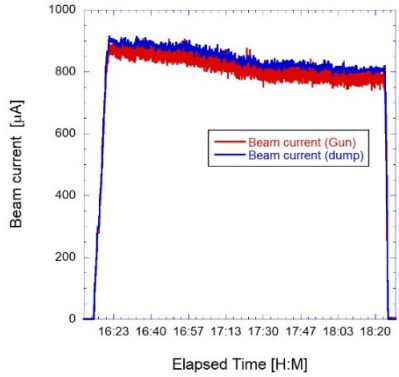

Figure 8: Beam current trend during energy recovery at approximately 1 mA CW operation in Run 2. The solid red line shows the beam current measured at the DC gun. The blue line shows the measured beam current at the beam dump by Faraday-cup. The horizontal (vertical) axis shows the elapsed time (beam current).

Fig. 8 shows the history of the beam current during the energy recovery of an approximately 1 mA CW beam in Run 2. We observed fluctuations in the measured beam current at the gun at approximately 1 mA, as shown in Fig. 8. Considering the measurement errors of both beam currents, the beam currents of the injection beam and dump section were transported with almost no loss, as shown in Fig. 8. The correlation between the laser and beam intensities in Run 2 is shown. Therefore, one of the beam drift sources shown in Fig. 8 may be the laser variation of this DC gun. No HOM heating was observed in the main-linac cavities. The vacuum level of the beam dump was maintained at less than $10^{-5}$ Pa without significant heating in the dump section by rastering the beam, whereas the nominal vacuum level without beam irradiation of the dump was at the $10^{-7}$ Pa level. The beam drifted from 0.9 mA to 0.8 mA within a few hours. However, we maintained stable beam operation. Fig. 9 shows the trend of the measured radiation by all ALOKA radiation monitors in the cERL during energy recovery at approximately 1 mA CW operation in Run 2, as shown in Fig. 8. ALOKA radiation monitors No. 5 and No. 6 were located immediately before and after the main-linac cryomodule, respectively [17]. Both monitors showed significant field emissions of approximately 100 mSv/h, which came from the main linac. The measured radiation in the No.5 and No.6 monitors in Fig. 9 is plotted by subtracting the background level, which originates from the main-linac field emission. These field emission levels did not change when the accelerating voltage did not change. However, the large background of the field emission might have enhanced the fluctuation of radiation monitors No. 5 and No. 6, as shown in Fig. 9. After collimator tuning, we maintained a low-radiation condition below 30 μSv/h in the cERL concrete shield. The drifts of the measured radiation of many ALOKA radiation monitors are shown as beam current drifting in Fig. 8. During this energy recovery operation, we did not observe a sudden increase in radiation due to a sudden beam kick or sudden beam instabilities, such as the HOM-BBU. Notably, we maintained stable beam operation in Runs 1 and 3 under approximately 1 mA CW beam energy recovery, even though we had the same drift for a few hours.

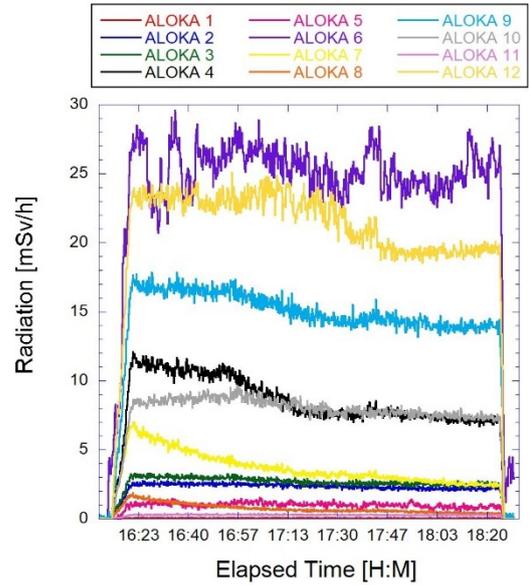

Figure 9: Trend of all ALOKA radiation monitors during energy recovery at approximately 1 mA CW operation in Run 2, as shown in Fig. 8.

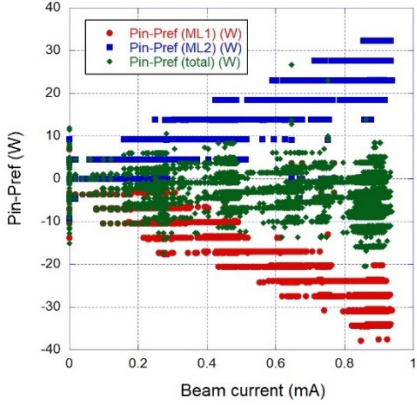

Figure 10: Power variation ($P_{in}$-$P_{ref}$) in Run 1 with respect to the beam current. The horizontal axis shows the beam current (I). The vertical axis shows the variation of $P_{in}$-$P_{ref}$ from the zero-beam current. The variations of $P_{in,cav}$-$P_{ref,cav}$ in each cavity (ML1 and ML2) are shown in red and blue, respectively. The variation of $P_{in}$-$P_{ref}$ for the sum of ML1 and ML2 is shown in green.

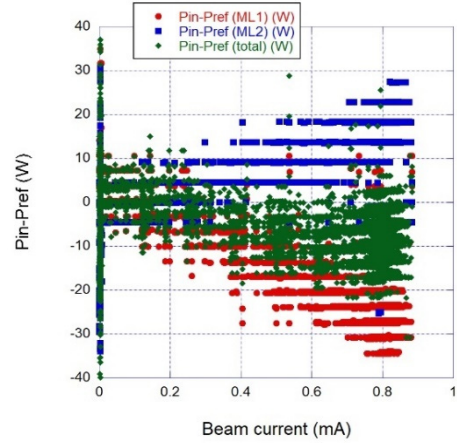

Figure 12: Power variation ($P_{in}$-$P_{ref}$) in Run 3 with respect to the beam current. The horizontal axis shows the beam current (I). The vertical axis shows the variation of $P_{in}$-$P_{ref}$ from the zero-beam current. The variation of $P_{in,cav}$-$P_{ref,cav}$ in each cavity (ML1 and ML2) are shown in red and blue, respectively. The variation of $P_{in}$-$P_{ref}$ for the sum of ML1 and ML2 is shown in green.

Figs. 10, 11, and 12 show the $P_{in}$-$P_{ref}$ variation during energy recovery in Runs 1, 2, and 3, respectively. The red and blue lines in these three figures show the difference in ($P_{in,cav}$-$P_{ref,cav}$) of each cavity with respect to the beam current in ML1 and ML2, respectively. The green line in these three figures shows the variation of ($P_{in}$-$P_{ref}$) with respect to the total ($P_{in}$-$P_{ref}$) of ML1 + ML2. We observed good linear responses between ($P_{in}$-$P_{ref}$) and the current I in Figs. 10, 11, and 12. $\varepsilon_{rf}$ was evaluated using a linear fit between ($P_{in}$-$P_{ref}$) and the beam current at each Run.

Table 3: Measured energy recovery efficiency ($\varepsilon_{rf}$).

| Beam condition | Total energy recovery efficiency ($\varepsilon_{rf}$) |
|---|---|
| Run 1 | 100.032% ± 0.031% |
| Run 2 | 99.958% ± 0.035% |
| Run 3 | 100.040% ± 0.035% |

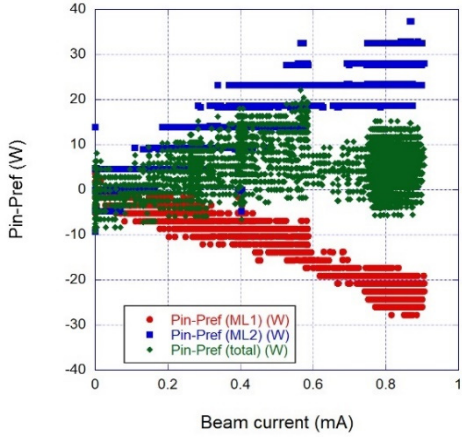

Figure 11: Power variation ($P_{in}$-$P_{ref}$) in Run 2 with respect to the beam current. The horizontal axis shows the beam current (I). The vertical axis shows the variation of $P_{in}$-$P_{ref}$ from zero-beam current. The variations of $P_{in,cav}$-$P_{ref,cav}$ in each cavity (ML1 and ML2) are shown in red and blue, respectively. The variation of $P_{in}$-$P_{ref}$ for the sum of ML1 and ML2 is shown in green.

Table 3 lists the energy recovery efficiencies in Figs. 10, 11, and 12 for Runs 1, 2, and 3, respectively. Although there was a deviation of less than 0.042% in Runs 1, 2, and 3, the energy recovery rate was almost 100.00%, considering an error bar of 0.035%. The measured error bars in Table 3 come from the accuracy of the power meter and the error of linearity of the power meters of $P_{in}$ and $P_{ref}$ (see details in Appendix A). Notably, we stably operated with amplitude stability of less than 0.02% and phase stability of less than 0.02° for ML1 and ML2 during beam operation [22].

# IV DISCUSSION

## IV-1 Beam loss discussion

The accelerator in the cERL is surrounded by a concrete shield. If large beam loss occurs, the generated photons reach the exterior of the concrete shield near the beam-loss point. Therefore, the beam loss distribution along the beamline in the cERL with beam loss will reflect the photon dose rate distribution on the roof of the cERL concrete shield. A high-current operation was achieved by suppressing this beam loss in the cERL. Therefore, we roughly determined the beam loss location by measuring the photon dose rates on the concrete roof and estimated the beam loss current during an energy recovery operation of approximately 1 mA. Fig. 13 shows the measured photon dose rate distribution for Run 1. A 1-inch NaI(Tl) scintillation survey meter (Aloka, TCS-171B) was used for these measurements. The maximum value of 0.23 µSv/h was measured on the concrete roof on the north straight line.

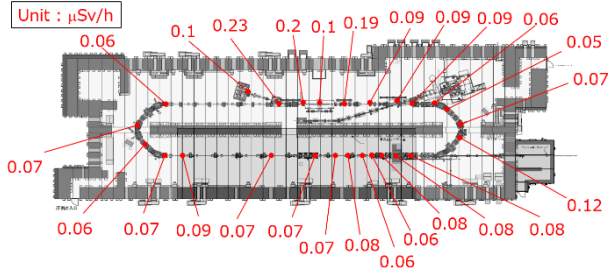

Figure 13: The measured photon dose rate distribution on the concrete roof on Run 1. The photon dose rate values include a background value of 0.05 µSv/h. The beam current was between 0.8 and 0.9 mA.

The photon dose rate per beam loss current was calculated using the MARS 15 code to estimate the beam loss current corresponding to the measured photon dose rate on the concrete roof [23,24]. In the case of beam loss at the electric magnet, 20 MeV electrons were bombarded with the beam duct on the inner surface upward by 1° in the center of the electric magnet. For beam loss at the collimator, 20 MeV electrons were bombarded with the collimator in the beam direction. An example of the MARS 15 calculation results is shown in Fig. 14. The maximum photon dose rate appeared slightly downstream from the beam-loss point, as shown in Fig. 14. In this case, a beam loss of 1 nA leads to a photon dose rate of 0.027 µSv/h on the roof of the cERL room downstream from the chicane magnets. The calculation results of the 17.5 MeV beam are similar to the results of the 20 MeV beam. Therefore, we applied the simulation results to the 17.5- and 20.0-MeV cases.

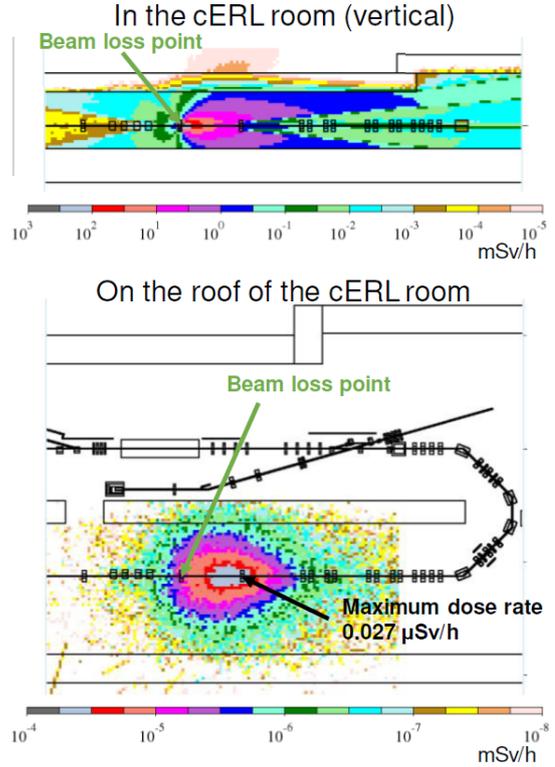

Figure 14: Example of the dose rate distribution calculated using MARS15. In this calculation, 1 nA beam loss occurred in the center of the electric magnet [23].

For Run 1, the maximum measured radiation was 0.23 µSv/h on the roof downstream of the main linac, as shown in Fig. 13. If we apply the beam loss estimation, as shown in Fig. 14, near the maximum radiation point of 0.23 µSv/h with the same thickness of the roof at this point, the beam loss current can be roughly estimated to 0.23 µSv/h / 0.027 µSv/h x 1 nA = 0.009 µA at this point. The loss ratio was found to be 0.009 µA/0.9 mA = 0.001%. The energy recovery efficiency was 100.032% ± 0.031% in Run 1. We did not compare the actual beam loss estimates because the local loss point was not measured in Run 1. However, the total loss was estimated by $(1 - \varepsilon_{rf})$, and the loss estimation of 0.001% by the simulation of MARS15 from the measured maximum values of 0.23 µSv/h, as shown in Fig. 13, was estimated to be nearly consistent with -0.032% ± 0.031% of $(1 - \varepsilon_{rf})$. Many dose meters must be set along the cERL beamline to estimate the detailed beam loss distribution.

Other energy losses occurred when the beam current increased. With respect to the HOM heat load during the approximately 1 mA operation, the HOM heat load is calculated using our design value of 10 V/pC with a 3-ps bunch length of the loss factor of the HOM in the main linac [18]. In Runs 1 and 2, the calculated HOM heat load was 6.3 mW and in Run 3, the calculated HOM heat load was 54 mW. These HOM heat loads contributed little to the energy recovery efficiency. These heat loads resulted in no HOM heating during the energy recovery operation of approximately 1 mA.

We summarized that an energy recovery efficiency of almost 100.00% can be achieved by eliminating losses at all locations along the beam line in the cERL when energy recovery is successful, as demonstrated in Run 1.

*IV-2 Emittance measurement discussion*

In Runs 1 and 2, high-efficient energy recovery operation of approximately 1 mA was achieved with very small emittance of 0.2–0.4 mm mrad. These values are also close to the measurement results shown in Ref. [14], indicating that the space-charge effect is negligible, even when the charge increases to 0.7 pC/bunch. Conversely, the higher emittances of approximately 1–2 mm mrad were obtained in Run 3, where energy recovery was performed at 5.5 pC/bunch. These emittance growths are mainly due to space-charge effects [25]. In particular, the calculated emittance was 2.9 mm mrad in the horizontal direction and 1.8 mm mrad in the vertical direction when the emittance was calculated in the case of Run 3 optics with a beam of 7.7 pC/bunch at the south straight section. These values are similar to the measured values in Table 2 for Run 3. The energy recovery operation was possible with an emittance as small as a few mm mrad, which is close to the design of Run 3. For 10-mA energy recovery operations in the future, cERL can be operated at 7.7 pC/bunch at 1.3 GHz. If the beam optics are the same as in Run 3 for a 10-mA energy recovery operation, energy recovery without beam loss will be possible.

## V CONCLUSION

This study investigates the energy recovery operations with a cERL at approximately 1 mA. Three energy recovery runs were performed under three different conditions. We performed beam tuning with low emittance and extremely small beam loss before CW operation by optimum beam tuning and by effectively using the collimator and loss monitor. The measured emittances were as small as 0.2–0.4 mm mrad at 0.7 pC/bunch. Furthermore, even at 5.5 pC/bunch operation, a low emittance of 2 mm mrad was achieved. After switching to CW operation, extremely high-efficient energy recovery operation of 100.0% ± 0.04% were achieved in three energy recovery runs at an approximately 1 mA beam. This 100.0% level of energy recovery reduced the beam loss and achieved very low radiation conditions.

This method, which efficiently uses collimators and loss monitors, enables energy recovery efficiency of 100.0% at a high current of 1 mA with low emittance and extremely small beam loss. In particular, the realization of energy recovery at 5.5 pC/bunch shows the feasibility of future 10 mA high current operations with an extremely high energy recovery efficiency. The findings of this study will provide insight into steady beam operation for high-brightness beams and high-efficiency energy recovery operations required for future EUV-FELs [6] and other applications.

## ACKNOWLEDGEMENTS

We would like to thank Hiroshi Kawata, Yukinori Kobayashi, and Shinichiro Michizono for their continuous encouragement of the cERL beam operation. We thank the cERL members for their support with the beam operation. This work was partially supported by the Government (MEXT) Subsidy for Strengthening Nuclear Security and the Quantum Beam Technology Program of the Japanese Ministry of Education, Culture, Sports, Science, and Technology (MEXT).

## APPENDIX

*A Consideration of energy recovery of each cavity on main linac*

In the cERL, ($P_{in,cav}$-$P_{ref,cav}$) of ML1 and ML2 had small slopes with respect to beam current, respectively, as shown in Figs. 10, 11, and 12. If the beam energy is sufficiently high, the beam velocity is close to the speed of light. In this case, energy recovery is perfectly established, and no current dependence appears. The total injection energy at a cERL of 2.9 MeV is extremely low, and the beam does not reach the speed of light under acceleration by the main linac. This causes a phase slip during acceleration and deceleration, resulting in slightly different efficiencies during acceleration and deceleration for each cavity. We performed a particle-tracking simulation in a simple mode to confirm whether the measured nonenergy recovery rate was reasonable.

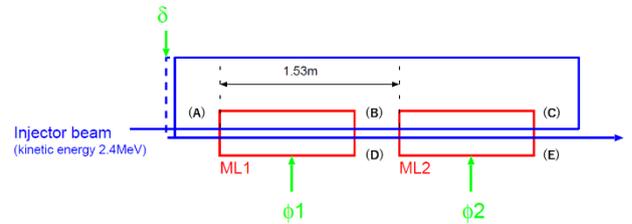

Figure A.1: The calculation setup layout of cERL under energy recovery conditions.

Fig. A.1 shows the setup for calculations. A beam with an injection energy of 2.9 MeV (kinetic energy of 2.4 MeV) was injected into the main linac, which was based on two 1.3 GHz nine-cell cavities of the ERL-model-2 cavity (ML1, ML2) [17,18]. Fig. A.2 shows the electrical field profile of the main linac along the beam axis, defined as Ez. The distance between the centers of the upstream (ML1) and downstream cavities (ML2) was 1.53 m. The amplitude and phase of each cavity were controlled independently. After acceleration in the two cavities, the beam was given a delay in the recirculation loop, injected into ML1, and tracked to the exit of ML2. The phase of the beam after the orbit was adjusted using the delay length, and the phase relationship between the two cavities was fixed.

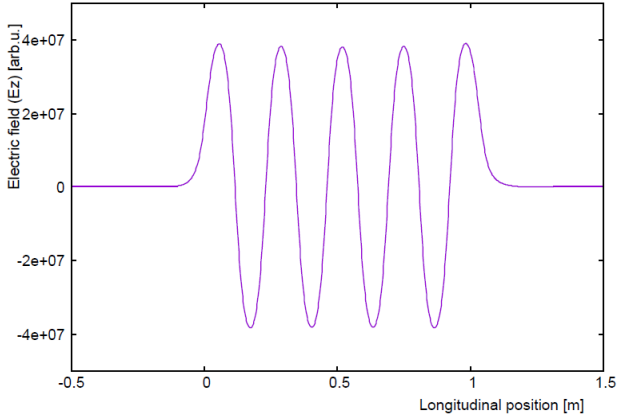

Figure A.2: The electrical field profile of the cERL main linac. The horizontal (vertical) axis shows the longitudinal position along the beam axis in the cavity (the electrical field of the beam direction (arbitrary unit)), respectively.

The calculation setup was similar to that for the cERL accelerator. The procedure is as follows: (1) The phase of ML1 ($\phi1$) is scanned and set to the phase in which the energy is maximum at the ML1 exit during acceleration (on-crest). (2) The ML2 phase ($\phi2$) is scanned and set to the phase where the energy is maximum at the ML2 exit during acceleration (on-crest). (3) The "delay" of the recirculation loop ($\delta$) is adjusted and set so that the energy at the ML2 exit is minimized after deceleration without changing the phases of $\phi1$ and $\phi2$. (4) The beam energy is calculated and the kinetic energy at each point is obtained (Fig. A.1 (A)–(E)) after the phase and delay length adjusted.

Table A.1: Energy gain calculation of each cavity on the main linac under energy recovery condition at cERL.

| Position | Kinetic energy (MeV) | Gain of each cavity (MV) |
|---|---|---|
| (A) | 2.4000 | |
| (B) | 10.8805 | 8.4805 (ML1) |
| (C) | 19.4034 | 8.5229 (ML2) |
| (D) | 10.8797 | -8.5238 (ML1) |
| (E) | 2.4003 | -8.4794 (ML2) |

The kinetic energy of the cERL injection beam was 2.4 MeV, and that of the beam after acceleration on the linac was 19.4 MeV. A total of 17.0 MeV was obtained in the main linac and accelerated in the main linac cavities. Table A.1 shows the calculated energies at positions (A)–(E) for ML1 and ML2 with an 8.5-MV amplitude. The accelerating energy of ML1 was obtained from the energy difference between (A) and (B): The accelerating energy of ML2 was obtained from the energy difference between (B) and (C): The energy difference between (C) and (D) provides the deceleration energy of ML1. The energy difference between positions (A) and (E) provides the deceleration energy of ML2.

The energy recovery efficiency is defined in Eq. (2). By contrast, the inefficiency of energy recovery ($\varepsilon_{rf\_ineff}$) is defined using Eq. (3) for a clearer comparison.

$\varepsilon_{rf\_ineff}$ (%) = (1- $\varepsilon_{rf}$)(%) = ($P_{in}$-$P_{ref}$)/ $P_{rf,acc}$ × 100%  (3)

Eq. (3) defines the energy recovery inefficiency of cERL operation. When we estimate the inefficiency of the energy recovery of each cavity of ML1 and ML2, we can replace $P_{in}$ ($P_{ref}$) to $P_{in,cav}$ ($P_{ref,cav}$) of each cavity of Eq. (3) and $P_{rf,acc}$ to the beam power of each cavity as $V_c$ × I. From the balance of the cavity acceleration under the energy recovery condition, we can redefine the inefficiency of the energy recovery as follows:

$\varepsilon_{rf\_ineff}$ (%) = ($V_{c\_acc}$ + $V_{c\_dec}$)/(($V_{c\_acc}$ − $V_{c\_dec}$)/2) × 100%, (4)

where $V_{c\_acc}$ ($V_{c\_dec}$) is the acceleration (deceleration) gain of each cavity. In other words, for ML1, $V_{c\_acc}$ is the gain obtained from (A) to (B), and $V_{c\_dec}$ is the gain obtained from (C) to (D). For ML2, $V_{c\_acc}$ is the gain obtained from (B) to (C), and $V_{c\_dec}$ is the gain obtained from (D) to (E).

A comparison of the calculated and measured energy inefficiencies of energy recovery for each cavity and the total is shown in Table A.2. The measured values in Table A.2 were used in Run 1 and Run 3, as shown in Figs. 12 and 13, respectively, and the inefficiency of the energy recovery ($\varepsilon_{rf\_ineff}$) was obtained from the ($P_{in}$-$P_{ref}$) variation at the maximum beam current and ($P_{in}$-$P_{ref}$) at zero current, as shown in Figs. 10 and 12.

Table A.2: Comparison between the measured energy recovery inefficiency in Run 1 and Run 3 and the calculated energy recovery inefficiency.

| Energy recovery inefficiency ($\varepsilon_{rf\_ineff}$) | Experimental results (total energy of 20 MeV) | | Calculation |
|---|---|---|---|
| | Run 1 | Run 3 | |
| ML1 | -0.391% ± 0.047% | -0.453% ± 0.054% | -0.509% |
| ML2 | +0.326% ± 0.039% | +0.374% ± 0.044% | +0.512% |
| Total | -0.032% ± 0.031% | -0.040% ± 0.035% | +0.0018% |

The measurement results showed energy recovery inefficiencies of 0.3%–0.5% were found in each cavity. Conversely, the calculations show inefficiencies of approximately 0.51%. One of the measurements of the inefficiency of ML1 in Run 1 agrees well with the calculation within the error bar. However, the remaining measurement results for the inefficiencies in each cavity are slightly lower than the expected values. Some unknown systematic factors or errors remain under the cERL condition. For example, the fringe fields from ML1 and ML2 were larger, and the calculated inefficiency was larger. However, the measurement results were not significantly different from the calculation results. Therefore, approximately 100.0% energy recovery was achieved in the measurement and calculation results.

# REFERENCES

[1] M. Tigner, Nuovo Cimento **37**, 1228 (1965).

[2] W. Kaabi *et al.*, "PERLE: A High Power Energy Recovery Facility," in *Proc. IPAC2019*, Melbourne, Australia, May 2019, pp. 1396–1399. doi:10.18429/JACoW-IPAC2019-TUPGW008.

[3] F. Hug *et al.*, "MESA – An ERL Project for Particle Physics Experiments," in *Proc. LINAC2016*, East Lansing, MI, USA, September 2016, pp. 313–315. doi:10.18429/JACoW-LINAC2016-MOP106012.

[4] A. Bartnik *et al.*, "CBETA: First Multipass Superconducting Linear Accelerator with Energy Recovery," Phys. Rev. Lett. **125**, 044803 (2020). doi:10.1103/PhysRevLett.125.044803.

[5] K.-J. Kim *et al.*, "A Proposal for an X-Ray Free-Electron Laser Oscillator with an Energy-Recovery Linac," Phys. Rev. Lett. **100**, 244802 (2008). doi:10.1103/PhysRevLett.100.244802.

[6] H. Kawata, N. Nakamura, H. Sakai, R. Kato, and R. Hajima, "High power light source for future extreme ultraviolet lithography based on energy-recovery linac free-electron laser," J. Micro/Nanopattern. Mater. Metrol. **21**, 021210 (2022). doi:10.1117/1.JMM.21.2.021210.

[7] T. Akagi, A. Kosuge, S. Araki, R. Hajima, Y. Honda, T. Miyajima, M. Mori, R. Nagai, N. Nakamura, M. Shimada, T. Shizuma, N. Terunuma, and J. Urakawa, "Narrow-band photon beam via laser Compton scattering in an energy recovery linac," Phys. Rev. Accel. Beams **19**, 114701 (2016). doi:10.1103/PhysRevAccelBeams.19.114701.

[8] I. Drebot *et al.*, "BriXsinO High-flux dual x-ray and THz radiation source based on energy recovery linacs," in *Proc. IPAC2022*, Bangkok, Thailand, May 2022, pp. 2407–2410. doi:10.18429/JACoW-IPAC2022-THOXSP2.

[9] R. Hajima *et al.*, "First demonstration of energy-recovery operation in the JAERI superconducting linac for a high-power free-electron laser," Nucl. Instrum. Methods Phys. Res., Sect. A **507**, 115 (2003). doi:10.1016/S0168-9002(03)00849-0.

[10] G. R. Neil *et al.*, "The JLab high power ERL light source," Nucl. Instrum. Methods Phys. Res., Sect. A **557**, 9 (2006). doi:10.1016/j.nima.2005.10.047.

[11] A. Shevchenko *et al.*, "The Novosibirsk free electron laser facility," AIP Conf. Proc. **2299**, 020001 (2020). doi:10.1063/5.0031513.

[12] F. Schliessmann *et al.*, "Realization of a multi-turn energy recovery accelerator," Nat. Phys. **19**, 597 (2023). doi:10.1038/s41567-022-01856-w.

[13] B. Dunham *et al.*, "Record high-average current from a high-brightness photoinjector," Appl. Phys. Lett. **102**, 034105 (2013). doi:10.1063/1.4789395.

[14] M. Akemoto *et al.*, "Construction and commissioning of the compact energy-recovery linac at KEK," Nucl. Instrum. Methods Phys. Res., Sect. A **877**, 197 (2018). doi:10.1016/j.nima.2017.08.051.

[15] N. Nishimori, R. Nagai, S. Matsuba, R. Hajima, M. Yamamoto, T. Miyajima, Y. Honda, H. Iijima, M. Kuriki, and M. Kuwahara, "Generation of a 500-keV electron beam from a high voltage photoemission gun," Appl. Phys. Lett. **102**, 234103 (2013). doi:10.1063/1.4811158.

[16] K. Watanabe, S. Noguchi, E. Kako, K. Umemori, and T. Shishido, "Development of the superconducting rf 2-cell cavity for cERL injector at KEK," Nucl. Instrum. Methods Phys. Res., Sect. A **714**, 67 (2013). doi:10.1016/j.nima.2013.02.035.

[17] H. Sakai, E. Cenni, K. Enami, T. Furuya, M. Sawamura, K. Shinoe, and K. Umemori, "Field emission studies in vertical test and during cryomodule operation using precise x-ray mapping system," Phys. Rev. Accel. Beams **22**, 022002 (2019). doi:10.1103/PhysRevAccelBeams.22.022002.

[18] H. Sakai, K. Shinoe, T. Furuya, S. Sakanaka, T. Suwada, T. Takahashi, K. Umemori, and M. Sawamura, "Development of a 1.3 GHz 9-cell Superconducting cavity for the energy recovery linac," in *Proc. ERL07*, Daresbury, UK, 2007, pp. 56–61.

[19] T. Miyajima, Y. Honda, R. Takai, T. Obina, M.Shimada, K. Harada, M. Yamamoto, K. Umemori, H. Sakai, T. Miura, F. Qiu, N. Nakamura, S. Sakanaka, N. Nishimori, R. Nagai, R. Hajima, and D. Lee, "Status of higher bunch charge operation in Compact ERL," in *Proc. IPAC2015*, Richmond, VA, USA, May 2015, pp. 1583–1586. doi:10.18429/JACoW-IPAC2015-TUPWA067.

[20] O. Tanaka *et al.*, "New halo formation mechanism at the KEK compact energy recovery linac," Phys. Rev. Accel. Beams **21**, 024202 (2018). doi:10.1103/PhysRevAccelBeams.21.024202.

[21] M. Arnold *et al.*, "ERL Operation of S-DALINAC," in *Proc. ERL'19*, Berlin, Germany, 2019, pp. 1–4. doi:10.18429/JACoW-ERL2019-MOCOXBS02.

[22] F. Qiu, T. Matsumoto, S. Michizono, and T. Miura, "Development of MicroTCA-based Low-level Radio Frequency Control Systems for cERL and STF," in *Proc. 12th Int. Workshop on Emerging Technologies and Scientific Facilities Controls (PCaPAC2018)*, Hsinchu, Taiwan, Oct. 2018, pp. 124–126. doi:10.18429/JACoW-PCaPAC2018-THCA2.

[23] H. Matsumura *et al.*, "Beam Loss Estimation by Measurement of Secondarily Produced Photons under High Average-current Operations of Compact ERL in KEK," in *Proc. IPAC2016*, Busan, Korea, May 2018, pp. 695–698. doi:10.18429/JACoW-IPAC2016-WEPOR020.

[24] N. V. Mokhov and S. I. Striganov, "MARS15 overview," in *Proc. Hadronic Shower Simulation Workshop*, Batavia, IL, USA, Sep. 2006, AIP Conf. Proc. **896**, 50–60 (2007).

[25] T. Miyajima *et al.*, "60 pC Bunch Charge Operation of the Compact ERL at KEK," in *Proc. IPAC2017*, Copenhagen, Denmark, May 2017, pp. 890-893. doi:10.18429/JACoW-IPAC2017-MOPVA019.